# TITLE PAGE

**Title**

A network approach to quantifying radiotherapy effect on cancer: radiosensitive gene group centrality

**Authors**

Yu-Xiang Yao[1], Zhi-Tong Bing[2,3,4], Liang Huang[1*], Zi-Gang Huang[5*], Ying-Cheng Lai[6, 7]

**Affiliations**

1. School of Physical Science and Technology, Lanzhou University, Lanzhou 730000, P. R. China.

2. Evidence Based Medicine Center, School of Basic Medical Science of Lanzhou University, Lanzhou 730000, P. R. China.

3. Key Laboratory of Evidence Based Medicine and Knowledge Translation of Gansu Province, Lanzhou 730000, P. R. China.

4. Department of Computational Physics, Institute of Modern Physics, Chinese Academy of Sciences, Lanzhou 730000, P. R. China.

5. The Key Laboratory of Biomedical Information Engineering of Ministry of Education, National Engineering Research Center of Health Care and Medical Devices, The Key Laboratory of Neuro-informatics & Rehabilitation Engineering of Ministry of Civil Affairs, and Institute of Health and Rehabilitation Science, School of Life Science and Technology, Xi'an Jiaotong University, Xi'an 710049, P. R. China

6. School of Electrical, Computer and Energy Engineering, Arizona State University, Tempe, AZ 85287, USA.

7. Department of Physics, Arizona State University, Tempe, Arizona 85287, USA.

**Corresponding Authors**

huangl@lzu.edu.cn (Liang Huang).

huangzg@xjtu.edu.cn (Zi-Gang Huang).

## ABSTRACT

Radiotherapy plays a vital role in cancer treatment, for which accurate prognosis is important for guiding sequential treatment and improving the curative effect for patients. An issue of great significance in radiotherapy is to assess tumor radiosensitivity for devising the optimal treatment strategy. Previous studies focused on gene expression in cells closely associated with radiosensitivity, but factors such as the response of a cancer patient to irradiation and the patient survival time are largely ignored. For clinical cancer treatment, a specific pre-treatment indicator taking into account cancer cell type and patient radiosensitivity is of great value but it has been missing. Here, we propose an effective indicator for radiosensitivity: radiosensitive gene group centrality (RSGGC), which characterizes the importance of the group of genes that are radiosensitive in the whole gene correlation network. We demonstrate, using both clinical patient data and experimental cancer cell lines, which RSGGC can provide a quantitative estimate of the effect of radiotherapy, with factors such as the patient survival time and the survived fraction of cancer cell lines under radiotherapy fully taken into account. Our main finding is that, for patients with a higher RSGGC score before radiotherapy, cancer treatment tends to be more effective. The RSGGC can have significant applications in clinical prognosis, serving as a key measure to classifying radiosensitive and radioresistant patients.

**MAIN TEXT**
**Introduction**

Radiotherapy has been an indispensable tool for treating cancer and controlling its growth, which is received by nearly 50% of the cancer patients[1]. A fundamental issue in radiotherapy is to assess the radiosensitivity (RS) of a cancer patient to enable decision making toward optimal treatment strategy[2]. In the modern era of precise medicine, gene signatures as a response predictor for radiotherapy and chemotherapy have been effective in the treatment of various cancers[3-6]. Previous studies established that a group of 31 genes from NCI-60 cancer cells are closely associated with radiotherapy[7]. However, due to the high complexity of the gene regulatory system, the intrinsic relationship between these genes has been unclear and it remains difficult to predict the effectiveness of radiotherapy for cancer patients.

Network science has the potential to provide powerful tools for analyzing a variety of natural and artificial complex systems[8-11], including those in life science and medicine. For example, metabolic pathways can be identified through analyzing the network structure of genes/proteins[12, 13], epidemic outbreak can be explained based on spreading dynamics on networks[14, 15], ecological stability associated with environment can be assessed using the network approach[16], designing genetic circuits can benefit from the principles from network science[17], and tipping points in complex mutualistic networks can be analyzed and predicted[18]. Recently, network science has been employed to advance research in cancer and oncology, providing unprecedented insights into physiological phenomena related to tumor growth[19, 20]with clinical applications[21, 22], clarifying the biochemical factors and signaling pathways during primary tumor development[23, 24], and leading to pathway-directed drug discovery[25]. The purpose of this paper is to exploit principles of complex network science to propose, analyze, and validate a geometric indicator to effectively and quantitatively characterize the impact of radiotherapy on cancer patients. This is the radiosensitive gene group centrality (RSGGC), which can be calculated through identifying both the relations among gene signatures and the efficiency of the radiotherapy treatment.

Our work is motivated by the considerations that, in spite of the completeness of the Human Genome Project[26], the detailed functions and relationship among the genes remain unclear, and the existing protein-protein interaction (PPI) networks are unable to capture the overall features of the regulation process. It is thus useful to concentrate on the intrinsic properties of the genes with a network dynamic topology. In particular, we consider the intergenic correlations among data samples of the same cluster or classification to obtain a network through Pearson's correlation matrix, which is also known as the gene co-expression matrix[27]. The matrix quantifies the gene-gene relationship across tissue samples. In network science, the centrality of a node quantifies its importance in the network based on the local connection strength to its neighbors[28-30]. Similarly, the centrality of a cluster of nodes is the sum of the centrality values over all the nodes in the cluster, which characterizes the importance of the focal cluster[31]. Based on this concept, we propose RSGGC as a novel indicator to quantitatively assess malignant tumor development and to provide prognosis for patients. Combining with the finding of work[7], we group the 31 radiosensitive genes to be our focal cluster to calculate the RSGGC, as shown schematically in **Fig. 1**.

Through analysis based on different data sets from patients and cancer cell lines, we find a strong correlation between RSGGC and the patient's clinical or physiological indicators, suggesting that RSGGC can be potentially be applied to analyzing the therapeutic effect of radiotherapy. Specifically, for clinical data, we collect the previously public datasets from Gene Expression Omnibus (GEO) and the Cancer Genome Atlas (TCGA) to calculate the correlation between the survival time of a patient and his/her RSGGC value from the clinical perspective. We then analyze the the experimental data of cancer cell lines from the existing literature[32]. A typical data set includes the survival fraction (SF) of parallel cell lines with different radiation dosages (2Gy, 5Gy, 8Gy) as labels of irradiation resistance for follow-up process. We calculate the RSGGC values for the corresponding NCI mRNA expression data. A statistical analysis shows that patients with a high RSGGC score have a longer survival time after radiotherapy and, consistently, cell lines with a higher RSGGC value have a smaller survival fraction after irradiation. Further, for different glioblastoma multiforme (GBM) cancer subtypes, our results indicate that the higher the RSGGC score is, the better therapeutic effect would be for patients receiving radiotherapy. Finally, based on the time sequencing data associated with irradiation, we compare the values of RSGGC before and after irradiation and find a sensitive response of RSGGC to irradiation, indicating the potential advantage of using RSGGC to assess the effect of radiotherapy in a quantitative manner. Our detailed and systematic analysis of RSGGC suggest that it can be used as a potential indicator of the effectiveness of radiotherapy to greatly facilitate decision making toward an optimal strategy for treating cancer.

**Material and Methods**
**Data collection**

The required mRNA/cDNA expression profile of a patient or a cancer cell line involved in radiotherapy or in irradiation experiment, respectively, is collected from the previous published datasets from Gene Expression Omnibus (GEO) and the Cancer Genome Atlas (TCGA), with detailed information listed in **Table I**.

Radiotherapy is the leading therapeutic strategy for patients suffering from GBM, as the surgical risk for this type of cancer is greater than others. GBM thus provides substantially more radiotherapy cases in the test datasets (GSE7696[33], GSE16011[34], and TCGA-GBM). In order to investigate the effect of radiotherapy on patients, we use a number of cervical cancer patients in the dataset GSE3578[35] that contains the sequencing changes of mRNA expression during therapy. We use NCI-60 cancer cell lines in a more homogeneous environment as the validation datasets: GSE32474[36] and GSE5949[37], which were widely used in exploring the underlying mechanisms of cancer and for developing drugs[38-40]. As an additional validation of the cancer cell lines, we study the integrated datasets: GSE59[41] and GSE7505[32], which record the abundance of cDNA of the NCI-60 cell lines before and after irradiation, respectively.

**Data preprocessing**

For reliable and meaningful statistical analysis, preprocessing of data is necessary. Briefly, we first select proper data through ranking and variance cutoff. We then apply either linear or logarithmic scales to the pertinent index, e.g., the survival time of patients (**Table Ⅱ**) or the survival fraction of cell lines (**Table Ⅲ**). In the following, we describe the preprocessing details for clinical data and cancer cell lines separately.

*Preprocessing of clinical data.* For patients, we screen out the cases GSE7696, GSE16011 and TCGA-GBM from GEO and TCGA primarily. All clinical data are filtered according to the following criteria: (1) null and void cases were removed (for example, lost record, abnormal expression, and rewritten data), and patients with both mRNA expression and the corresponding clinical information were selected; (2) one clinical index was used to classify patients (e.g., survival time or subtypes of patients). The GSE3578 (cervical cancer patient set) has relatively adequate, complete clinical and gene expression information, so it was chosen as an additional validation of the ability of RSGGC to evaluate therapy, from which samples were classified according to the difference in therapeutic strategies and checking time. For convenience, samples were classified on a logarithmical scale to balance the survival time and the relevant order of magnitude of samples. **Table II** shows the physiological or pathological brief summary and grouping results.

*Preprocessing of cell lines data.* We match the description and annotation of the experimental samples with the expression profile, and remove samples with missing information. We adopt only the cancer cell lines described in previous work[32] and label them with the SF value. The radiosensitive and radioresistant lines can be represented by the replication rate, cell migration capacity, and SF under specific condition or environment. We use the SF value after receiving irradiation of different dosage level to characterize cell's radiosensitivity / radioresistance. All the cell line datasets are ranked according to the SF values of the cell lines after receiving 2Gy, 5Gy, and 8Gy irradiation, denoted by SF2, SF5, and SF8, respectively[32]. **Table III** shows the method of sampling groups for different radiation dosage, where the low dose irradiation dataset (2Gy) is grouped linearly according to SF2, such as deciles or quartiles, while the high dose irradiation datasets (5Gy and 8Gy) have logarithmical bins according to SF5 and SF8. The different ways of data binning were adopted just for convenience.

**Correlation matrix and RSGGC measure**

*Correlation matrix.* A basic fact in systems biology and biomedical science is that genes are not isolated with each other but work collectively as an interacting network, regardless of whether the underlying process is intracellular or extracellular. To characterize the responses of patients or cancel cell lines to irradiation, we introduce the measure of RSGGC by considering the differences among sample groups and the corresponding inherent dynamic relationship from the point of view of a complex network.

A prerequisite to defining RSGGC is the correlation matrix. To begin, for a given dataset, we rank all the genes by the inter-sample variance with a proper cutoff to ensure computational efficiency, taking into account the balance of heterogeneity of the platforms as in previous work[42]. The resulting M×N expression matrix (for M genes and N samples) are ordered again by some index, e.g., the survival rate or subtypes. We then divide the matrix into different sections: $M \times n_1, M \times n_2, \cdots$, where $n_1 + n_2 + \cdots = N$. For each sectional expression matrix, we calculate its Pearson's correlation matrix from all the gene pairs based on the available samples, which leads to an M-dimensional, fully connected, real symmetric matrix, with each element characterizing the similarity in the expression level of the two genes. To distinguish this M×M matrix from a gene co-expression network, we do not set any threshold so as to maintain the original correlation between the genes, and obtain an adjacency matrix without information loss, with the element of the matrix given by

$$\rho_{X,Y} = \frac{\mathrm{cov}(X,Y)}{\sigma_X \cdot \sigma_Y} \quad (1)$$

where $\mathrm{cov}(X,Y)$ stands the covariance between variables X and Y (genes), $\sigma_X$ and $\sigma_Y$ are the standard deviations (SD) of the two variables, respectively.

After the data is filtered and grouped, we calculate the intra-sample variance of a single gene for one total dataset. Some datasets provide gene's expression data, while others record the original expression profiles of RNA fragments. For the former, it is straightforward to calculate the variance but for the latter, we first merge the multiple probes that match the same genes via arithmetic averaging before calculating the variance. Probes without the corresponding gene names are dropped from the calculation. The detailed information of the remaining gene number in each step is presented in **Table S1**.

To treat the different datasets on an equal footing, it is necessary to determine a variance cutoff. In particular, the datasets are obtained from different platforms and are processed by probe merging, and each single dataset contains a different number of genes. One difficulty is that, after ranking by the variance size, many genes have values of variance that are close to each other. Moreover, invalid genes in the data lead to wasted computation. In addition, certain genes have almost the same expression among diverse cells or environment. To overcome these difficulties, we use the insights from previous work[43, 44] and choose the first 8000 genes with large variance to ensure that they are computationally distinct without loss of generality or universality of the results. Since not all 31 radiation sensitive genes are included in the top-8000 large variance genes, we generate the 31 genes contained in the gene list then complement it with the 8000th genes by means of variance. The variance distribution of the 31 genes is presented in **Table S2**).

*RSGGC measure.* In network science, a large number of centrality measures have been introduced in different contexts[45]. For example, degree centrality (DC) represents the number of edges of a node in the network, closeness centrality (CC) characterizes how close a pointed node from other nodes[46], betweenness centrality (BC) reflects a node's intermediary status of route or pathway in the network[47, 48], and eigenvector centrality (EC) measures the relative influence of a node in the network[49]. We focus on the EC of the correlation matrix as it is appropriate to our task. In general, the eigenvector $\vec{x}$ associated with the maximum eigenvalue of the adjacency matrix is closely related to the asymptotic behavior of the collective dynamics on the network and control[31]. Let $x_i$ be the component of the eigenvector corresponding to node $i$. The EC of node $i$ is given by

$$x_i = \frac{1}{\lambda} \sum_{j \epsilon N_i} x_j = \frac{1}{\lambda} \sum_{j \epsilon M} a_{ij} x_j \quad (2)$$

where $N_i$ is the set of node i's neighbors, M is the set of all nodes in the network, and $a_{ij}$ is the element of the Pearson's correlation matrix **A**. The vector form of Eq. 2 is

$$\mathbf{A} \cdot \vec{x} = \lambda \vec{x} \quad (3)$$

Our RSGGC measure is defined in terms of EC. In particular, for all the

selected genes, RSGGC is the ratio of the EC of the focal group to that of the entire gene set:

$$\text{RSGGC} = \frac{\sum_n |x_i|}{\sum_N |x_i|} \quad (4)$$

where $n$ is the number of genes in the radiation sensitivity gene group and N=8000 is the total number of genes in the whole set. For the available data set in our study, we have $n$=31 and N=8000. The absolute values in the sums indicate a focus on the importance of individual genes rather than distinguishing the detailed passive or active role of a specific gene in the system. The RSGGC value characterizes the topological and dynamical properties of a small group of genes in the whole gene network.

For a group, the correlation matrix represents its average level under various circumstances. The 8000×N matrices, with N being the sample size of each entire dataset, are treated as described in **Table Ⅱ** and **Table Ⅲ** so that the within-group RSGGC values can be computed. The multi-step analysis is summarized as a workflow chart, as shown in **Fig. 2**.

## Results
### Power of RSGGC as a predictor of radiotherapy outcome for clinical patients

We calculate RSGGC for each classified clinical patient group, as shown in **Fig. 3**. The striking finding is the robust positive correlation between the RSGGC value and the survival time, indicating that the patients with a higher RSGGC score have longer expected survival time after radiotherapy. That is, radiotherapy is more effective for patients with a higher RSGGC score. More specifically, the Pearson's correlation coefficient for the clinical data in **Fig. 3**(A-C) are 0.96, 0.83, and 0.87, respectively, with the significance index values of 0.06, 0.02, and 0.19 (**Table S3**). After random grouping of the clinical data, the positive correlation is lost completely (**Fig. S1**), providing strong evidence for the reliability of RSGGC as a quantitative indicator of the effect of radiotherapy. The finding of the positive correlation between RSGGC and patient survival time is unprecedented and practically significant, as it can serve as the base for more reliable prediction of the outcome of radiotherapy for cancer patients.

### Ability of RSGGC to predict radiation outcome in cell line experiments

Observation of the survival fraction of cancer cell lines in response to radiation in experiments is more straightforward and more controllable than clinical tests with patients. Our data analysis suggests that the results from experimental cell lines at the microscopic level strongly corroborate the role of RSGGC in clinical tests. **Fig. 4** shows the downtrend relationship between the RSGGC value and the survival fraction (SF), which is completely consistent with the results from the clinical data. In particular, the group of cancer cell lines with higher RSGGC scores has lower survival fraction after radiation, corresponding to longer survival time of patients.

The values of Pearson's correlation coefficients of RSGGC and SF in **Fig. 4** are -0.77 and -0.91, respectively, with the significance index values of 0.01 and 0.09 (More statistics are presented in Table S1). Similar results of negative correlation between RSGGC and SF have been obtained from the two datasets with larger irradiation doses (5Gy and 8Gy, see **Fig. S2**). In clinical practice, the general protocol for radiotherapy consists of daily exposure to fractionized radiation of 2Gy irradiation for 5~7 weeks. A dosage over 2Gy is in fact harmful to patient's health. Thus, the relationship between the survival fraction with 2Gy and RSGGC is practically significant for generating quantitative patient prognosis.

**Further exploitation of RSGGC for GBM subtypes**

The results in above focus on the relationship between RSGGC and the survival time for clinical cases of GBM, where information about the detailed subtypes of GBM is ignored. In a previous experimental study[50], it was found that the subtypes can have quite different radiation therapy effects. For example, radiation can have a significant effect on the subtypes "Classical", "Mesenchymal", and "Neural", but the effect is small for the subtype "Proneural". We ask whether RSGGC is capable of characterizing the radiation therapy effects at the subtypes level of GBM. To address this issue, we group cases according to the subtypes rather than the survival time and calculate the RSGGC values. **Table Ⅳ** presents the detailed clinical index and RSGGC for the four subtypes: Classical, Mesenchymal, Neural, and Proneural. We see that the first three subtypes have similar values of RSGGC, which are larger than that of the fourth subtype (Proneural). This coincides well with the therapeutic effect observed from the experimental studies of these subtypes. The general result is that radiation therapy is more effective for subtypes with a larger value of RSGGC.

**Mechanism of RSGGC as an estimator and predictor of radiotherapy effect**

Our computations and analysis taking into account the survival time of clinical patients, the survival fraction of cancer cell lines, or different subtypes, give strong evidence that RSGGC is effective for assessing and predicting the radiotherapy effect against cancer. A plausible reason for the power of RSGGC is that the corresponding selected genes may participate in the key pathways associated with repairing DNA damage, activating cell cycle checkpoints or maintaining signal transduction pathways after the irradiation, either directly or indirectly. To verify this conjecture, we employ clinical data[35] from cervical cancer patients prior to and during radiotherapy to test how the RSGGC scores calculated from the group of genes change as radiotherapy treatment is being implemented. Specifically, we classify the expression profile into two groups: prior to or during treatment, as shown in **Fig. 5**. We find that the RSGGC scores of patients receiving radiotherapy increase dramatically in comparison with those prior to the therapy. We also utilize the time sequences of samples with irradiation process obtained from two independent datasets: GSE59 and GSE7505, the cDNA microarrays of NCI-60 cell lines before and after irradiation[32, 41]. Since the datasets are from different experimental platforms, we consider only the common genes of the cell lines. After classifying the lines as describe in Materials and Methods, we calculate the RSGGC values before and after irradiation, as shown in **Fig. 5**. We find that, for the three different irradiation dosages, the RSGGC values of the groups with low SF decrease drastically while those with high SF increase, indicating that the effects -of irradiation at the molecular level vary for different cancer cell lines. The observable response of RSGGC to irradiation implies again its potential power in predicting the effect of irradiation therapy. In a general sense, RSGGC can effectively be regarded as a geometrical indicator of the activity of the radiosensitive gene group during irradiation. Since the centrality of the whole set of genes is normalized, an increase in the RSGGC score for a subset of genes implies that the centrality values of the remaining genes must decrease.

**Discussion**

A standard and widely used method to treat cancer patients is radiotherapy. An outstanding problem in medical science is to predict the survival time of a patient who has undergone radiotherapy. A major deficiency of previous work is the focus on gene expression in cells directly pertinent to radiosensitivity with factors such as the response of a cancer patient to irradiation and the patient survival time totally ignored. To overcome the deficiency and to devise a more accurate and reliable predictor of the patient survival time, we exploit modern complex network science to articulate a geometric approach to estimating and predicting the effect of radiotherapy on cancer. In particular, we propose a measure, the radiosensitive gene group centrality (RSGGC), that can be used to predict the survival time of a patient undergoing radiotherapy. We validate the predictive power of RSGGC by using data from both clinical patients and experimental cancer cell lines. Results from clinical data reveal a positive correlation between RSGGC and the survival time of the patients going through radiotherapy. Since, in clinical practice, a patient's prognosis is influenced by multiple factors[51, 52] that can introduce fluctuations in the outcomes (Table S1), we also systematically analyze data from cancer cell lines, which are more reliable due to the homogeneous microenvironment and the controllability of external conditions in experiments. Results from cancer cell lines support our finding from the clinical data in a completely consistent way: the cell lines with higher RSGGC values are more sensitive to irradiation and thus have smaller values of the survival fraction.

RSGGC as a novel indicator/predictor for characterizing radiosensitivity from a geometric viewpoint has potential advantages over the traditional clinical indicators. RSGGC can lead to new insights into understanding the relationship among the known radiosensitive gene signatures and can be used for data based analysis of extensive risk gene sets, intracellular pathways regulation and control.

**References**

[1] Moding EJ, Kastan MB, Kirsch DG. Strategies for optimizing the response of cancer and normal tissues to radiation. Nat Rev Drug Disc. 2013;12:526-42.

[2] Eschrich S, Zhang H, Zhao H, Boulware D, Lee J-H, Bloom G, et al. Systems biology modeling of the radiation sensitivity network: a biomarker discovery platform. Inter J Radia Oncol Biol Phys. 2009;75:497-505.

[3] Zhu C-Q, Ding K, Strumpf D, Weir BA, Meyerson M, Pennell N, et al. Prognostic and predictive gene signature for adjuvant chemotherapy in resected non--small-cell lung cancer. J Clin Oncol. 2010;28:4417-24.

[4] Van De Vijver MJ, He YD, van't Veer LJ, Dai H, Hart AAM, Voskuil DW, et al. A gene-expression signature as a predictor of survival in breast cancer. New Engl J Med. 2002;347:1999-2009.

[5] Lee HJ, Lee J-J, Song IH, Park IA, Kang J, Yu JH, et al. Prognostic and predictive value of NanoString-based immune-related gene signatures in a neoadjuvant setting of triple-negative breast cancer: relationship to tumor-infiltrating lymphocytes. Breast Cancer Res Treat. 2015;151:619-27.

[6] Bing Z, Tian J, Zhang J, Li X, Wang X, Yang K. An Integrative Model of miRNA and mRNA Expression Signature for Patients of Breast Invasive

Carcinoma with Radiotherapy Prognosis. Cancer Biothera Radiopharma. 2016;31:253-60.

[7] Kim HS, Kim SC, Kim SJ, Park CH, Jeung H-C, Kim YB, et al. Identification of a radiosensitivity signature using integrative metaanalysis of published microarray data for NCI-60 cancer cells. BMC Geno. 2012;13

[8] Palla G, Der\'e nI, Farkas Ies, Vicsek Tas. Uncovering the overlapping community structure of complex networks in nature and society. Nature. 2005;435:814-8.

[9] Barthe lM. Spatial networks. Phys Rep. 2011;499:1-101.

[10] Holme P, Saram\"a kJ. Temporal networks. Phys Rep. 2012;519:97-125.

[11] Wang W-X, Lai Y-C, Grebogi C. Data based identification and prediction of nonlinear and complex dynamical systems. Phys Rep. 2016;644:1-76.

[12] Kauffman S, Peterson C, Samuelsson Bor, Troein C. Random Boolean network models and the yeast transcriptional network. Proc Nat Acad Sci (USA). 2003;100 14796-9

[13] Barabasi A-L, Oltvai ZN. Network biology: understanding the cell's functional organization. Nat Rev Genet. 2004;5:101-13.

[14] Pastor-Satorras R, Vespignani A. Epidemic spreading in scale-free networks. Phys Rev Lett. 2001;86:3200.

[15] Brockmann D, Helbing D. The hidden geometry of complex, network-driven contagion phenomena. Science. 2013;342:1337-42.

[16] Proulx SR, Promislow DEL, Phillips PC. Network thinking in ecology and evolution. Trends Ecol Evolu. 2005;20 345-53

[17] Alon U. Network motifs: theory and experimental approaches. Nat Rev Genet. 2007;8 %6:450-61.

[18] Jiang J, Huang Z-G, Seager TP, Lin W, Grebogi C, Hastings A, et al. Predicting tipping points in mutualistic networks through dimension reduction. Proc Natl Acad Sci (USA). 2018:201714958.

[19] Yarden Y, Pines G. The ERBB network: at last, cancer therapy meets systems biology. Nat Rev Cancer. 2012;12:553-63.

[20] Pujana MA, Han J-DJ, Starita LM, Stevens KN, Tewari M, Ahn JS, et al. Network modeling links breast cancer susceptibility and centrosome dysfunction. Nat Genet. 2007;39:1338-49.

[21] Chuang H-Y, Lee E, Liu Y-T, Lee D, Ideker T. Network-based classification of breast cancer metastasis. Mole Sys Biol. 2007;3:140.

[22] Gevaert O, De Smet F, Timmerman D, Moreau Y, De Moor B. Predicting the prognosis of breast cancer by integrating clinical and microarray data with Bayesian networks. Bioinfo. 2006;22:e184-e90.

[23] Ivliev AE, Ac't Hoen P, Sergeeva MG. Coexpression network analysis identifies transcriptional modules related to proastrocytic differentiation and sprouty signaling in glioma. Cancer Res. 2010;70:10060-70.

[24] Balkwill F. Cancer and the chemokine network. Nat Rev Cancer. 2004;4:540-50.

[25] Altieri DC. Survivin, cancer networks and pathway-directed drug discovery. Nat Rev Cancer. 2008;8:61-70

[26] Venter JC, Adams MD, Myers EW, Li PW, Mural RJ, Sutton GG, et al. The sequence of the human genome. Science. 2001;291:1304-51.

[27] Stuart JM, Segal E, Koller D, Kim SK. A gene-coexpression network for global discovery of conserved genetic modules. Science. 2003;302:249-55.

[28] Freeman LC. Centrality in social networks conceptual clarification.

Soc Net. 1978;1:215-39.

[29] Gomez D, Gonz\'a l-AueE, Manuel C, Owen G, del Pozo M, Tejada J. Centrality and power in social networks: a game theoretic approach. Math Soc Sci. 2003;46:27-54.

[30] Borgatti SP, Everett MG. A graph-theoretic perspective on centrality. Soc Net. 2006;28:466-84.

[31] Aguirre J, Papo D, Buld\'u JM. Successful strategies for competing networks. Nat Phys. 2013;9:230-4.

[32] Amundson SA, Do KT, Vinikoor LC, Lee RA, Koch-Paiz CA, Ahn J, et al. Integrating global gene expression and radiation survival parameters across the 60 cell lines of the National Cancer Institute Anticancer Drug Screen. Cancer Res. 2008;68:415-24.

[33] Murat A, Migliavacca E, Gorlia T, Lambiv WL, Shay T, Hamou M-F, et al. Stem cell--related "self-renewal" signature and high epidermal growth factor receptor expression associated with resistance to concomitant chemoradiotherapy in glioblastoma. J Clin Oncol. 2008;26:3015-24.

[34] Gravendeel LAM, Kouwenhoven MCM, Gevaert O, de Rooi JJ,

Stubbs AP, Duijm JE, et al. Intrinsic gene expression profiles of gliomas are a better predictor of survival than histology. Cancer Res. 2009;69:9065-72

[35] Iwakawa M, Ohno T, Imadome K, Nakawatari M, Ishikawa K-i, Sakai M, et al. The radiation-induced cell-death signaling pathway is activated by concurrent use of cisplatin in sequential biopsy specimens from patients with cervical cancer. Cancer Biol Thera. 2007;6:905-11.

[36] Pfister TD, Reinhold WC, Agama K, Gupta S, Khin SA, Kinders RJ, et al. Topoisomerase I levels in the NCI-60 cancer cell line panel determined by validated ELISA and microarray analysis and correlation with indenoisoquinoline sensitivity. Mole Cancer Therap. 2009;8:1878-84.

[37] Reinhold WC, Reimers MA, Lorenzi P, Ho J, Shankavaram UT, Ziegler MS, et al. Multifactorial regulation of E-cadherin expression: an integrative study. Mole Cancer Thrap. 2010:1535-7163.

[38] Shoemaker RH. The NCI60 human tumour cell line anticancer drug screen. Nat Rev Cancer. 2006;6:813-23.

[39] Gao H, Korn JM, Ferretti Sep, Monahan JE, Wang Y, Singh M, et al. High-throughput screening using patient-derived tumor xenografts to

predict clinical trial drug response. Nat Med. 2015;21:1318-25.

[40] Lee JK, Havaleshko DM, Cho H, Weinstein JN, Kaldjian EP, Karpovich J, et al. A strategy for predicting the chemosensitivity of human cancers and its application to drug discovery. Proc Nat Acad Sci (USA). 2007;104:13086-91.

[41] Ross DT, Scherf U, Eisen MB, Perou CM, Rees C, Spellman P, et al. Systematic variation in gene expression patterns in human cancer cell lines. Nat Genet. 2000;24:227-35.

[42] Volinia S, Croce CM. Prognostic microRNA/mRNA signature from the integrated analysis of patients with invasive breast cancer. Proc Nat Acad Sci (USA). 2013;110:7413-7.

[43] Oldham MC, Horvath S, Geschwind DH. Conservation and evolution of gene coexpression networks in human and chimpanzee brains. Proc, Nat Acad Sci (USA). 2006;103 17973-8

[44] Zhang B, Horvath S. A general framework for weighted gene co-expression network analysis. Stat Appl Gene Mole Biol. 2004;4:Article17.

[45] Newman MEJ. Networks: An Introduction: Oxford University Press; 2010.

[46] Bavelas A. Communication patterns in task-oriented groups. J Acou Soc Ame. 1950;22:725-30.

[47] Freeman LC. A set of measures of centrality based on betweenness. Sociometry. 1977:35-41.

[48] Park KH, Lai YC, Ye N. Characterization of weighted complex networks. Phys Rev E. 2004;70:026109.

[49] Bonacich P. Power and centrality: A family of measures. Ame J Socio. 1987;92:1170-82.

[50] Verhaak RGW, Hoadley KA, Purdom E, Wang V, Qi Y, Wilkerson MD, et al. Integrated genomic analysis identifies clinically relevant subtypes of glioblastoma characterized by abnormalities in PDGFRA, IDH1, EGFR, and NF1. Cancer Cell. 2010;17:98-110.

[51] Berchuck A, Iversen ES, Lancaster JM, Pittman J, Luo J, Lee P, et al. Patterns of gene expression that characterize long-term survival in advanced stage serous ovarian cancers. Clin Cancer Res. 2005;11:3686.

[52] Wang MJ, Ping J, Li Y, Adell G, Arbman G, Nodin B, et al. The prognostic factors and multiple biomarkers in young patients with colorectal cancer. Sci Rep. 2015;5:10645.

**Figures & Figure Legends**

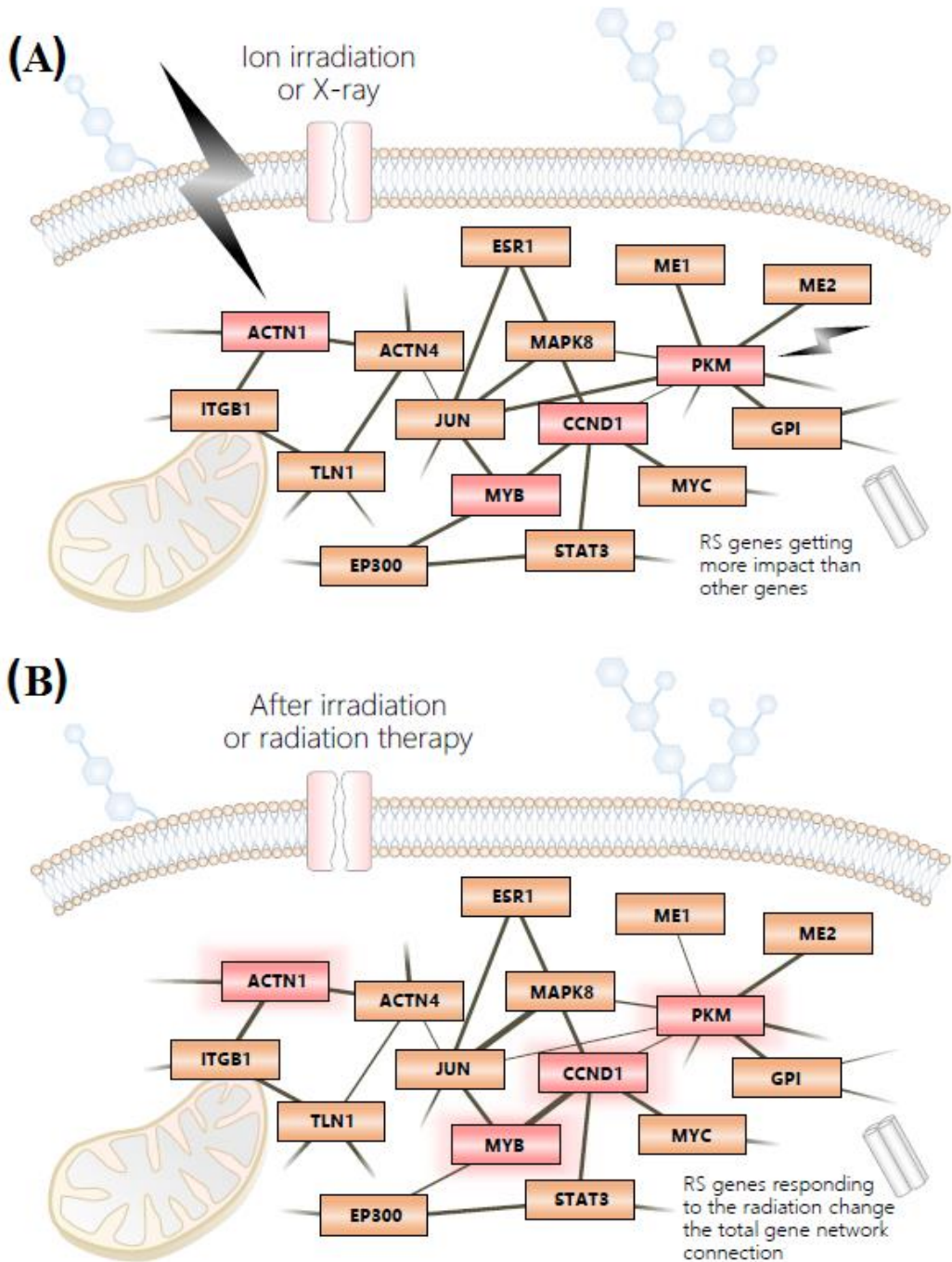

**Figure 1. Schematic illustration of the effects of radiation sensitive genes on the intercellular interaction or correlation of genes in cells:** (A) before and (B) after irradiation. The red and orange blocks represent the radiation sensitive and the remaining genes, respectively. For simplicity

and clarity, only a small fraction of the genes are demonstrated. The links in the network are drawn according to the open database STRING-DB. The thickness of a line between a pair of genes indicates the interaction strength between the genes. A comparison between (A) and (B) gives a glimpse of one possible change in the interaction strength due to irradiation.

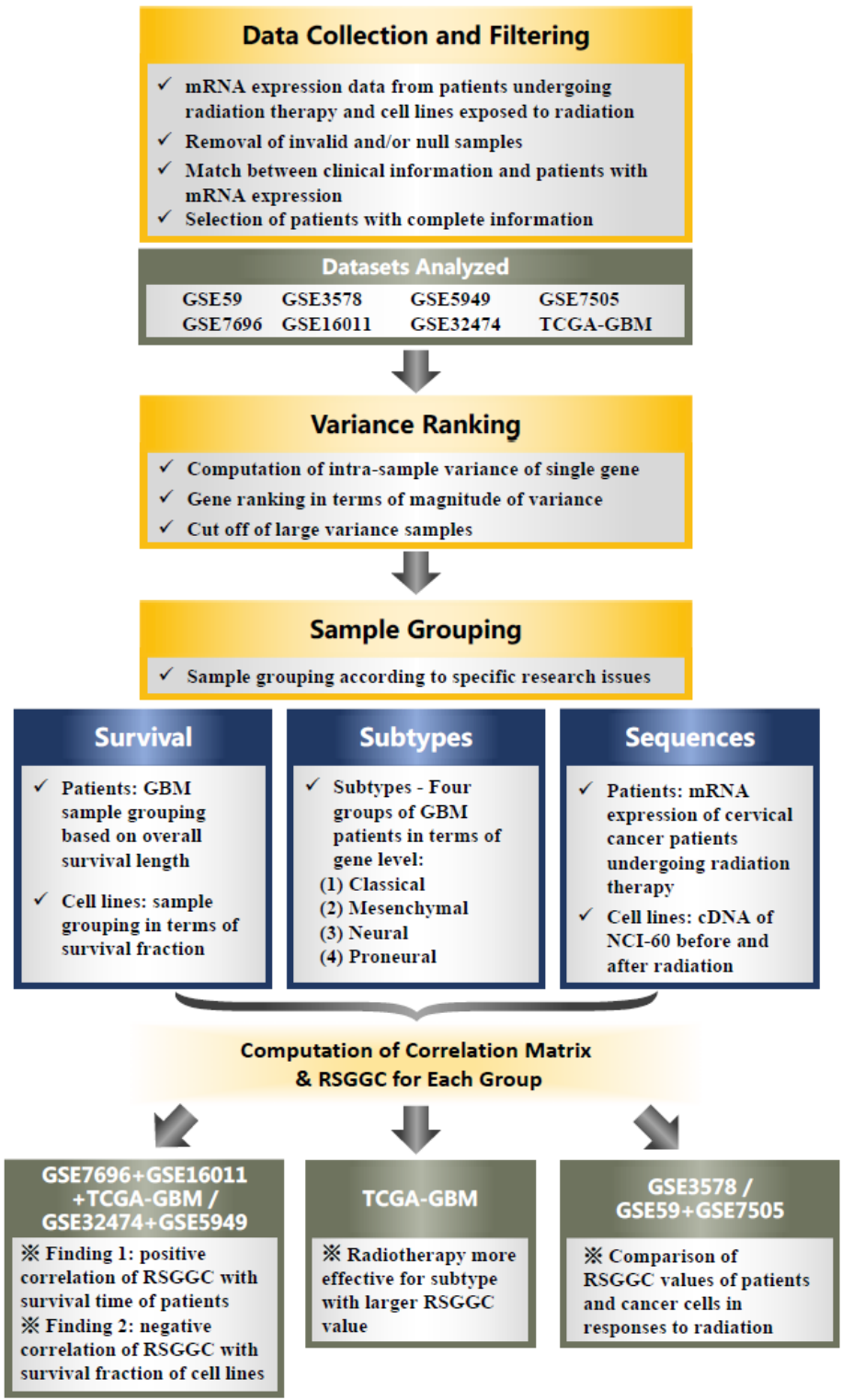

**Figure 2. Workflow of multi-step calculation of RSGGC.**

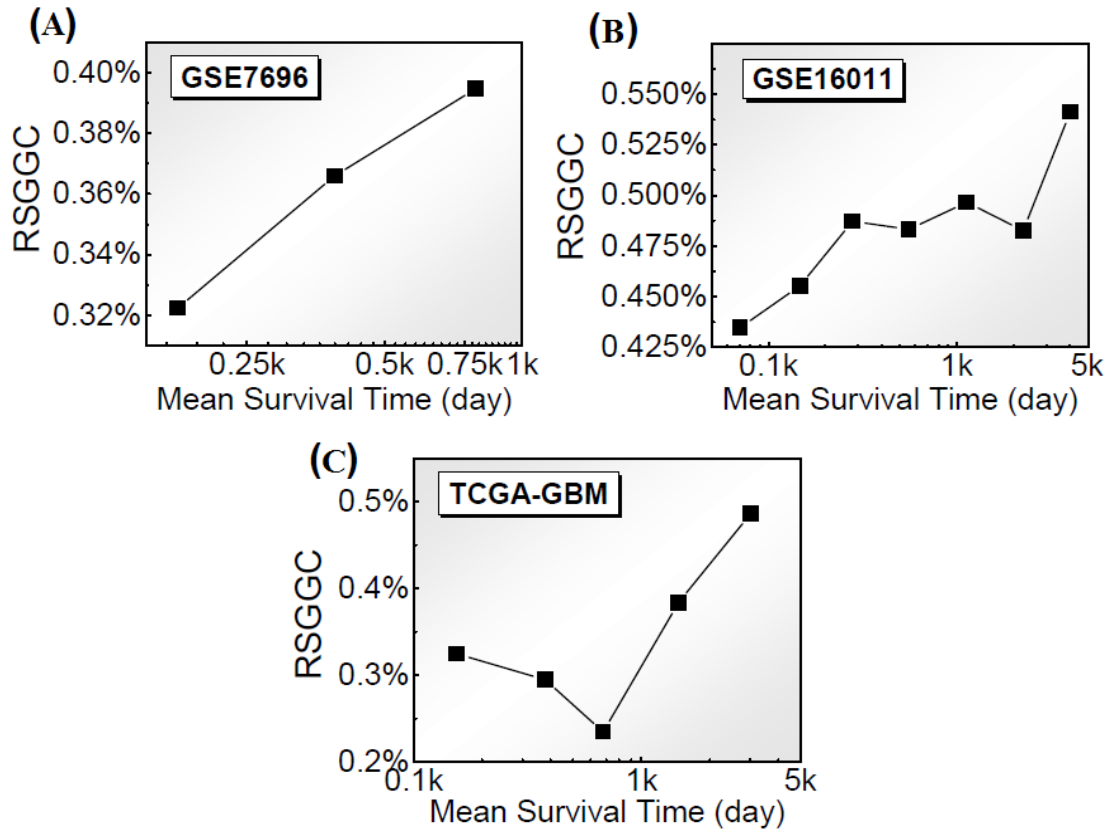

**Figure 3. Correlation between RSGGC score and survival time of GBM patients.** Datasets are (A) GSE7696, (B) GSE16011, and (C) TCGA-GBM. The horizontal axis denotes the mean survival time for each sample group (black squares). A semi-logarithmic scale is used. The positive correlation between patients' RSGGC value and the mean survival time has been identified in all the datasets. The dip in the RSGGC score in the region of small survival time in (C) is due to the unusually short-living samples in the dataset.

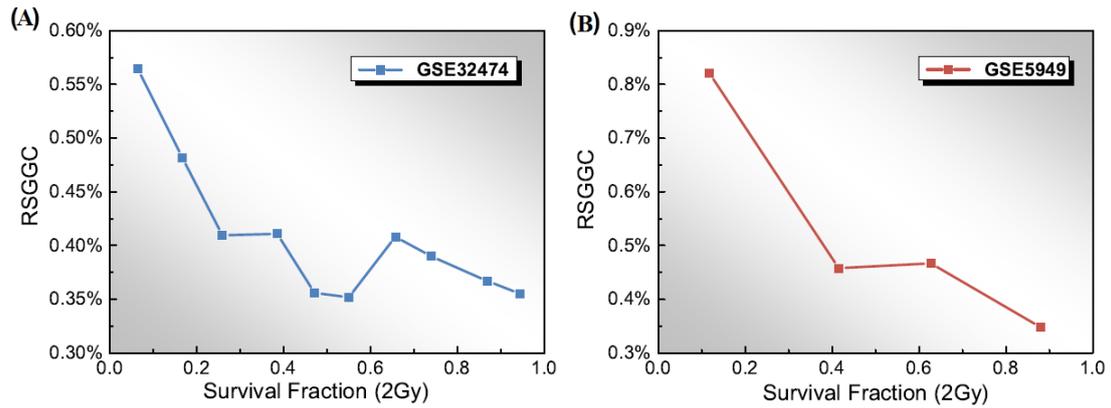

**Figure 4. Negative correlation between RSGGC and survival fraction of cancer cell lines.** The datasets are (A) GSE32474 and (B) GSE5949 with 2Gy radiation. The horizontal axis is the mean SF value of the samples within the group. RSGGC exhibits a negative relationship with the survival fraction, which is completely consistent with the results in Fig. 3.

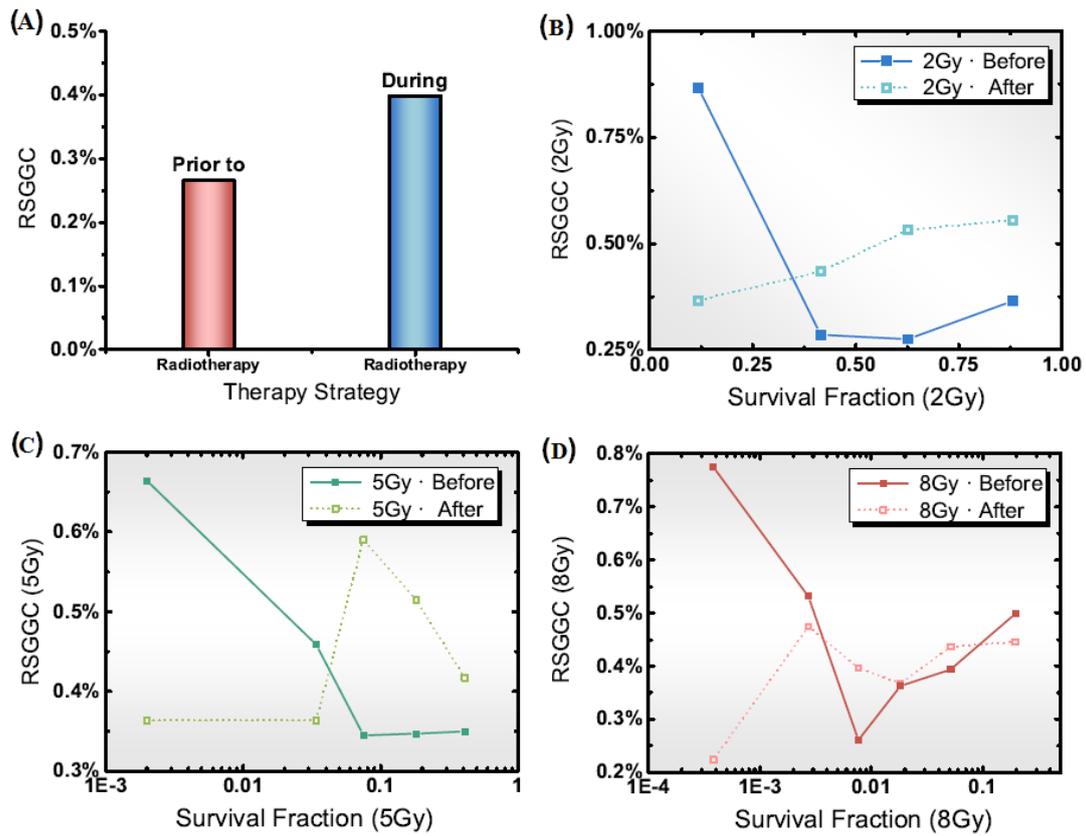

**Figure 5. Applicability of RSGGC to cervical cancer patients and cell lines.** (A) RSGGC values of cervical cancer patients prior to or during irradiation treatment. There is a dramatic increase in the RSGGC value as a result of the irradiation treatment. (B-D) RSGGC values of NCI-60 cancer cell lines before and after irradiation, with dosages of 2, 5, and 8Gy, respectively. The solid and open squares represent the RSGGC values of each group of cell lines before (from data in GSE59) and after receiving the irradiation (from data in GSE7505).

**Tables in Manuscript**

**TableⅠ. Datasets used in this article.**

| Dataset | Brief Description | Platform[1] |
|---|---|---|
| GSE59 | cDNA microarrays of NCI-60 cancer cell lines | GPL167, 169 |
| GSE3578 | mRNA expression of cervical cancer patients during therapy | GPL2895 |
| GSE5949 | mRNA expression of NCI-60 cancer cell lines | GPL91~95 |
| GSE7505 | cDNA microarrays of NCI-60 cancer cell lines after radiation | GPL5080 |
| GSE7696 | mRNA expression of GBM patients | GPL570 |
| GSE16011 | mRNA expression of GBM patients | GPL8542 |
| GSE32474 | mRNA expression of NCI-60 cancer cell lines | GPL570 |
| TCGA-GBM | GBM patient samples from TCGA | GPL570 |

[1] We adopt GEO accession for data sets to simplify the description of the GEO Platform (GPL). The detailed information, which can be found from the website of National Center for Biotechnology Information (NCBI), is:
   (1) GPL91~95, Affymetrix Human Genome U95A U95E Array;
   (2) GPL167, 169, 10kPrint3, 10kPrint2 of spotted DNA/cDNA (non-commercial);
   (3) GPL570, Affymetrix Human Genome U133 Plus 2.0 Array;
   (4) GPL2895, GE Healthcare/Amersham Biosciences CodeLink Human Whole Genome Bioarray;
   (5) GPL5080, NHGRI Homo sapiens 6K of spotted DNA/cDNA;
   (6) GPL8542, Affymetrix GeneChip Human Genome U133 Plus 2.0 Array.

**Table Ⅱ. Clinical Information and Groups of Patients in Different Datasets**

| Dataset | Sample Size | Grouping to the Survival Time (K days) | Mean Survival Time (K days)[1] | Mean Age | Gender (M/F) |
|---|---|---|---|---|---|
| GSE7696 | 65 | 0~0.25~0.50~∞ | 0.438 | 51.178 | 48/17 |
| GSE16011 | 180 | 0~0.1~0.2~0.4~ 0.8~1.6~3.2~∞ | 0.836 | 49.986 | 122/58 |
| TCGA-GBM | 318 | 0~0.25~0.50~ 1.00~2.50~∞ | 0.563 | 56.138 | 195/123 |

[1] Only the patients with a clinical event (death) are taken into account.

**Table Ⅲ. Method to group cancer cell lines based on the survival fractions under different radiation dosages (2, 5, and 8Gy, respectively).**

| Dataset | Sample Size | Grouping to the Survival Fraction | | |
|---|---|---|---|---|
| | | SF2 | SF5 | SF8 |
| GSE59 | 60 | quartiles | | |
| GSE5949 | 60 | quartiles | 0~0.01~0.05~ 0.10~0.25~1.00 | 0~0.001~0.005~ 0.010~0.25~1.00 |
| GSE7505 | 60 | quartiles | | |
| GSE32474 | 174 | deciles | 0~0.005~0.05~ 0.1~0.25~1.00 | 0~0.0005~0.005~0.010~ 0.025~0.05~0.1~1.0 |

**Table Ⅳ. Characteristic description of clinical subtypes of GBM and the corresponding RSGGC values.**

| Subtypes | Sample Size[1] | Age (mean) | Gender (M/F) | Therapeutic Effect[2] | HR[2] | P-value[2] | RSGGC(%) |
|---|---|---|---|---|---|---|---|
| Classical | 141 | 10.9~86.59 (58.9) | 83/58 | significant | 0.45 | 0.02 | 0.321 |
| Mesenchymal | 151 | 24.4~84.8 (59.6) | 91/60 | significant | 0.54 | 0.02 | 0.357 |
| Neural | 82 | 23.1~88.6 (59.9) | 54/28 | effective | 0.56 | 0.1 | 0.334 |
| Proneural | 129 | 17.7~89.3 (53.9) | 77/52 | less effective | 0.80 | 0.4 | 0.223 |

[1]  The total sample size is 503, which includes patients with clinical events (death).
[2]  The therapeutic effect, the values of hazard ratio (HR), and P-values are from Ref. [50].

Supplementary Materials for

# A network approach to quantifying radiotherapy effect on cancer: radiosensitive gene group centrality

Yu-Xiang Yao, Zhi-Tong Bing, Liang Huang, Zi-Gang Huang, and Ying-Cheng Lai
Corresponding authors: Liang Huang and Zi-Gang Huang

## 1. Genes filtrating & probes merging

As mentioned in the main text, the datasets used in this work were obtained from various platforms (**Table-1** in the main text). **Table-S1** describes the detailed filtrating or merging process for datasets.

**Table-S1 Data Filtrating and Merging Processes**

| Dataset | Platform[1] | Types | Genes or Fragments | Original Rows | Removing NULL or invalid* | Merging Multiple Probes | Utilizing in Paper | RS genes Included |
|---|---|---|---|---|---|---|---|---|
| GSE59 | GPL167, 169 | cDNA | fragments | 9984[2] | 7208 | 6182 | 1445[3] | 6 |
| GSE3578 | GPL2895 | mRNA | fragments | 54676 | 24467 | 19578 | 8000 | 30 |
| GSE5949 | GPL91~95 | mRNA | fragments | 63178[4] | 42499 | 19962 | 8000 | 31 |
| GSE7505 | GPL5080 | cDNA | fragments | 6728 | 6126 | 5743 | 1445[3] | 6 |
| GSE7696 | GPL570 | mRNA | fragments | 54675 | 45782 | 23519 | 8000 | 31 |
| GSE16011 | GPL8542 | mRNA | genes | 17527 | 17454 | 17454 | 8000 | 29 |
| GSE32474 | GPL570 | mRNA | fragments | 54675 | 45782 | 23519 | 8000 | 31 |
| TCGA-GBM | GPL570 | mRNA | genes | 12402 | 12402 | 12402 | 8000 | 31 |

[1] We use the GEO accession as in the main text.

[2] Only GPL167 is described.

[3] There 1445 remaining genes outside of the intersection between GSE59 and GSE7505 (See also **Table-S2**).

[4] This row shows the integrating results of the five platforms GPL91~GPL95, and the number of genes for each platform is 12626 (12182), 12621 (9118), 12647 (7541), 12645 (5535), and 12639 (8123).

## 2. Program Realization

We carried out data filtrating using software R (version 3.1.3 in this paper), MS Excel 2013 and MATLAB2014a (v8.3.0.532). The R program was applied to initial screening of the valid genes / fragments or to matching the expression with clinical data / survival fraction / surviving time / subtypes information. In addition, we merge the RNA fragments by R order `DATA2<-aggregate.data.frame(DATA1,by=list (DATA$genes), FUN=mean)`, where `DATA1` is the original mRNA fragments data, `DATA2` is the outcome of merging, and all data processed by arithmetic average according to gene names `DATA$genes`. The variance can be obtained by using R or Excel with the executive order being the `answer<-apply(DATA,1,var)`, or function `VAR.P( )`, respectively. Considering that MATLAB is specialized in matrix computation, we deal with grouped data through the function `corr( )` to get the correlation matrix and to obtain the maximum eigenvector by order `[v,~]=eig(matrix)`. It is noteworthy that not all 31 radiation sensitive genes are in the top 8000 (see **Table-S2**).

**Table-S2 Distribution of the variance of 31 RS genes from the corresponding datasets**

| 31 Genes | GSE59[1] | GSE3578 | GSE5949[2] | GSE7505 | GSE7696 | GSE16011 | GSE32474 | TCGA-GBM |
|---|---|---|---|---|---|---|---|---|
| ACTN1 | 1167/392 | 2469 | 2992 | 2501 | 1922 | 868 | 4723 | 1016 |
| ANXA2 | 1101/—— | 138 | 8865 | —— | 1587 | 614 | 2348 | 964 |
| ANXA5 | 1324/—— | 3845 | 8571 | 5569 | 1862 | 2273 | 3124 | 2929 |
| ARHGDIB | 94/—— | 686 | 47 | 2141 | 4921 | 2494 | 733 | 1653 |
| CAPNS1 | 4074/—— | 2779 | 17411 | 5223 | 4168 | 5160 | 7397 | 6294 |
| CBR1 | 853/—— | 815 | 2086 | 2186 | 2488 | 239 | 803 | 1165 |
| CCND1 | —— | —— | 1143 | 3763 | 1233 | 1069 | 458 | 796 |
| CD63 | 2871/—— | 616 | 6433 | —— | 2639 | 4193 | 3702 | 7035 |
| CORO1A | 2131/—— | 4775 | 117 | 4294 | 2713 | 2240 | 605 | 1363 |
| CXCR4 | —— | 2904 | 338 | —— | 898 | 707 | 280 | 778 |
| DAG1 | 1292/790 | 1920 | 8227 | 5544 | 11032 | 6025 | 3964 | 5565 |
| EMP2 | 1161/—— | 1271 | 4029 | 4696 | 2406 | 1623 | 1497 | 568 |
| HCLS1 | 874/—— | 8057 | 2933 | 1347 | 1069 | 574 | 263 | 669 |
| HTRA1 | 326/240 | 332 | 268 | —— | 1968 | 3075 | 196 | 861 |
| ITGB5 | 596/—— | 15919 | 1207 | 5355 | 5210 | 1282 | 3309 | 3316 |
| LAPTM5 | 543/61 | 587 | 569 | 3564 | 1332 | 992 | 374 | 824 |
| LRMP | 139/302 | 6361 | 4811 | 2045 | 8883 | 3072 | 2043 | 5590 |
| MYB | 4923/—— | 1886 | 2366 | 2936 | 16495 | 3944 | 742 | 3264 |
| PFN2 | 609/—— | 10789 | 271 | 4365 | 2061 | 4161 | 322 | 1507 |
| PIR | —— | 8636 | 406 | 5083 | 724 | 495 | 354 | 500 |
| PKM2[2] | 3995/—— | 137* | 14907* | —— | 4029* | 3424 | 6799* | 2708 |
| PTMS | 3334/—— | 3855 | 6979 | 5065 | 13411 | 1717 | 13837 | 8874 |
| PTPRC | 44/11 | 8529 | 5900 | 2706 | 5007 | 1571 | 1507 | 968 |
| PTPRCAP | 4847/—— | 17377 | 749 | 2630 | 13145 | 13894 | 2335 | 8606 |
| PYGB | 1385/—— | 5217 | 10898 | 4949 | 12048 | 5042 | 9241 | 2343 |
| RAB13 | —— | 546 | 1737 | —— | 5070 | —— | 970 | 3607 |
| RALB | 2531/—— | 7154 | 12357 | 4412 | 9849 | 7619 | 5060 | 4883 |
| SCRN1 | 456/3 | 16020 | 764 | —— | 2641 | 3386 | 329 | 1230 |
| SQSTM1 | 2839/1293 | 675 | 13261 | —— | 10817 | —— | 8074 | 3537 |
| TWF1 | —— | 10019 | 18719 | 1812 | 20938 | 10608 | 10242 | 5007 |
| WAS | 1014/272 | 3887 | 4125 | 2215 | 17564 | 14531 | 14110 | 9423 |
| **Total** | **26/10** | **30** | **31** | **23** | **31** | **29** | **31** | **31** |

[1] Two columns represent the variance distribution of two GPL167, GPL169 respectively.

[2] GPL91~GPL95 utilized in GSE5949 are analyzed integrally here.

[3] Line symbol means the platforms doesn't contain the gene; the star symbol means some datasets we choose PKM as the RS gene instead of PKM2.

## 3. Biostatistics information

**Table-S3** lists the detailed information about the biostatistics of the results in the main text.

**Table-S3 Biostatistics information of the results in the main text**

| Dataset | Pearson's Coef | $R^2$ | P-value | Group Num. |
|---|---|---|---|---|
| GSE32474(SF2)[1] | -0.76993 | 0.54189 | 0.00919 | 10 |
| GSE5949(SF2)[1] | -0.91284 | 0.7499 | 0.08716[2] | 4 |
| TCGA-GBM | 0.87008 | 0.67606 | 0.0551 | 5 |
| GSE16011 | 0.83134 | 0.62935 | 0.02044 | 7 |
| GSE7696 | 0.9569 | 0.83133 | 0.18758[2] | 3 |

[1] Included is only the 2Gy dose case.
[2] The P-values of GSE5949 and GSE7696 are large because of the small amount of grouping of the two datasets. The absolute values of Pearson's correlation coefficient of the two curves are close to one, implying underlying regularity.

## 4. Random grouping of three clinical datasets

We divide samples into random groups regardless of patient's survival time and calculate the corresponding RSGGC values. The black solid lines in Fig. S1 correspond to the results from the whole sample set without grouping. For the randomized cases, there is no indication of any correlation between the RSGGS value and the patients' survival time demonstrated in the main text.

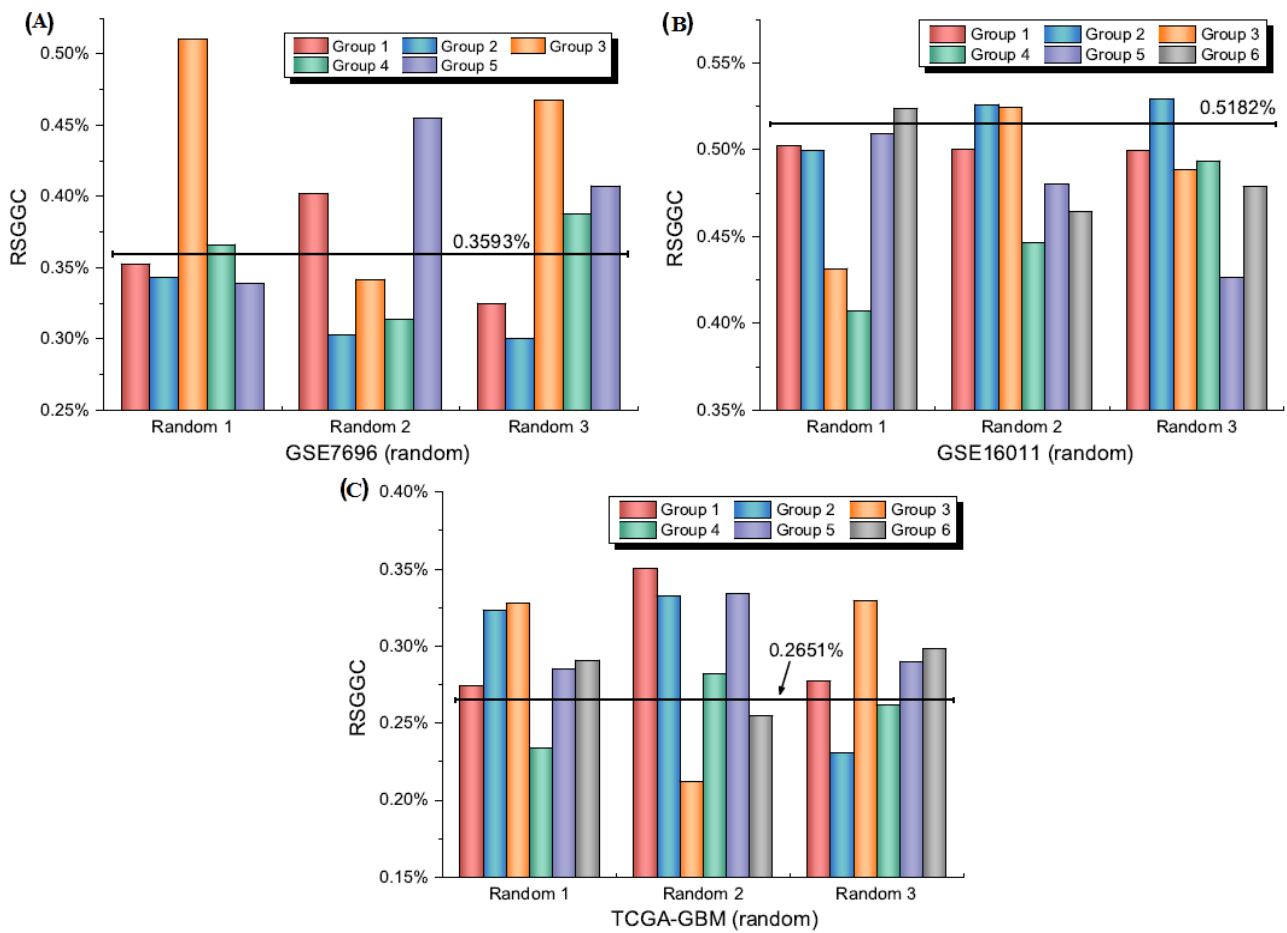

**Figure-S1 RSGGC values of random grouping of clinical datasets.**

## 5. RSGGC of cancer cell lines in high irradiation dose

The results from cancer cell lines under relatively high irradiation doses (5 Gy and 8 Gy) are shown in **Fig-S2** (the corresponding results for the case of 2 Gy can be found in **Fig. 4** in the main text).

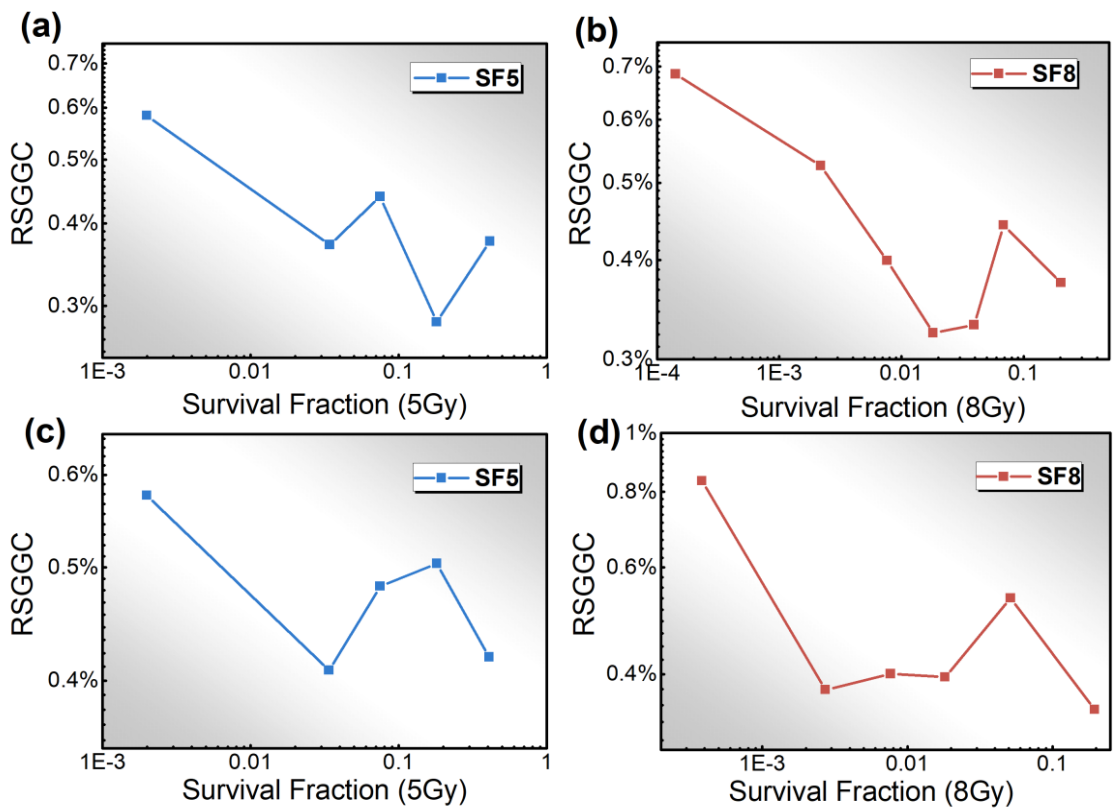

**Figure-S2 RSGGC values of cancer cell lines subject to higher irradiation dose.** The top and bottom rows show the results from GSE32474 and GSE5949, respectively. The left and right columns are results for irradiation doses of 5Gy and 8Gy, respectively.